\documentclass[useAMS,usenatbib]{mn2e}

\usepackage{graphicx}
\usepackage{fleqn}
\def\dfrac#1#2{{\displaystyle#1\over\displaystyle#2}}

\title{
Gravitational lensing by cosmic strings: what we learn from the
CSL-1 case. }
\author[Sazhin Mikhail]
{Sazhin M.V.$^1$,  Khovanskaya O.S.$^{1}$, Capaccioli
M.$^{2,3}$,Longo G.$^{3,4,5}$,
\newauthor
Paolillo M.$^{3,5,6}$, Covone G.$^{4}$, Grogin N.A.$^{7}$, Schreier E.J.$^{7,8}$ \\
 1 - Sternberg Astronomical Institute, Moscow State University, Universitetskii pr. 13, Moscow, RUSSIA\\
 2 - VSTceN-INAF - via Moiariello 16, 80131, Napoli ITALY\\
 3 - University of Napoli Federico II, Department of Physical Sciences, via Cinthia 6, 80126 Napoli, ITALY\\
 4 - INAF - Osservatorio Astronomico di Capodimonte, via Moiariello 16, Napoli, ITALY\\
 5 - INFN - Napoli Unit, via Cinthia 6, 80126, Napoli, ITALY \\
 6 - Space Telescope Science Institute, Baltimore - MD, USA\\
 7 - Dep. of Physics and Astronomy, The Johns Hopkins University, Baltimore, MD 21218, USA\\
 8 - Associated Universities Inc., Washington, DC 20036, USA\\
  }

\begin{document}

\date{Accepted ;
      Received ;
      in original form }
\pagerange{\pageref{firstpage}--\pageref{lastpage}} \pubyear{2005}

\maketitle

\label{firstpage}

\begin{abstract}
Cosmic strings were postulated by Kibble in 1976 and, from a
theoretical point of view, their existence finds support in modern
superstring theories, both in compactification models and in
theories with extended additional dimensions. Their eventual
discovery would lead to significant advances in both cosmology and
fundamental physics. One of the most effective ways to detect
cosmic strings is through their  lensing signatures
which appear to be significantly different from those introduced
by standard lenses (id est, compact clumps of matter). In 2003,
the discovery of the peculiar object CSL-1 \citep{csl1} raised the
interest of the physics community since its morphology
and spectral features strongly argued in favour of it being the first case of
gravitational lensing by a cosmic string. 
In this paper we provide a detailed description of the expected observational
effects of a cosmic string and show, by means of simulations, the lensing signatures
produced on background galaxies. While high angular resolution images obtained
with HST, revealed that CSL-1 is a pair of interacting ellipticals
at redshift $0.46$, it represents a useful lesson to plan future surveys.

\end{abstract}


\section{Introduction}\label{introduction}

Cosmic strings as topological defects of space-time were
introduced by \citet{kib76} and have been thoroughly discussed in
cosmology over the past decades \citep[cf.][]{zel, vil, string1}.
Among all possible types of such defects cosmic string are
preferable arising in inflation scenarios and find support in
modern theoretical physics. The great progress in cosmic string
theory has been achieved within superstring theories, both in
compactification models and in theories with extended additional
dimensions.

The main cosmic string parameter (i.e. the linear density $\mu$) 
depends strongly on the underlying model and may vary over a wide range, 
even though some constrains can be obtained from superstring theory
\citep{dav05, cop04, maj05, tye05}. However all cosmic strings, either
classical strings, or F- and D-strings, share two properties which
are model independent: the extremely long cosmological length and
a negligibly small cross-section.

Without doubts, identification of cosmic string parameters will
allow to distinguish the underlying theory. But first of all it is
necessary to answer the principal question: do cosmic
strings exist in our Universe?

From the observational point of view, the most evident signature
of a cosmic string is that it must induce gravitational lensing
effects on background sources  producing a strip ("milky way") of
multiple images along its path. However, theory predicts that
strings can be very far from the observer, thus requiring ultra
deep whole sky galaxy surveys to maximize the possibilities of
detection.

The second observational signature arises from the huge ratio
existing between the string width and length, which leads to a
sort of step function signature on the images of background
sources. As it has already been shown in \cite{csl1} and will be
further discussed in what follows, this implies that the lensing
of an extended objects by a cosmic string produces sharp edges in
the isophotes of the lensed object: a phenomenon which cannot be
found in standard gravitational lensing by compact objects. To
test this property, the angular resolution of the observations is
crucial since, as will be discussed in more detail in what
follows, the angular size of the lensing signatures is related to
the angular size of string strip.

Obviously the probability to observe such effects depends on the
expected number of cosmic strings. While most
estimates \citep{all, J.Polchinski, Bennet90, Rigv05} predict a few dozen long
strings crossing horizon volume, simulations using an
underlying field theory \citep{vin98, bev04, bev06} show that the
long string density can be significantly
lower (by about a factor 4) than suggested by earlier simulations,
and the loop density is negligible. In any case so far all attempts to detect the
expected gravitational signatures seem to have failed (see for
instance \cite{shi04}). In \cite{csl1}, and \cite{saz05} some of
us discussed the unusual properties of a peculiar extragalactic
object (hereafter CSL-1) which, by a careful analysis of its
photometric and spectroscopic investigation seemed to be a good
candidate. In fact, CSL-1 looks as a double source projected
against a low density field. The two components are separated by
$1.9$ arcsec, and result clearly extended even in ground
based optical images. Detailed photometry showed that both
components had identical shapes within the limits of ground based
images. Low and medium-high resolution spectra pointed out that
also the spectra of the two components were identical at a $98 \%$
confidence level, and gave a differential radial velocity of $0
\pm 20$ km s$^{-1}$ at a redshift of $z = 0.46 \pm 0.008$. These
observational evidences led to two possible explanations: either
CSL-1 was a rare close pair of two very similar and isolated giant
elliptical galaxies, or it was a gravitational lens phenomenon. In
the latter case, detailed modeling showed that the properties of
CSL-1 could be explained only by the lensing of an E-type galaxy
by a cosmic string.

In fact, the most relevant feature of the two CSL-1 images is that
their isophotes appeared to be undistorted down to the faintest
light levels, while the usual gravitational lenses (i.e. lenses
created by a bound clump of matter) produce inhomogeneous
gravitational fields which always distort the multiple images of
extended background sources \cite[cf.][]{sch92,kee}. As
pointed out in \cite{csl1}, one way to disentangle in a non
ambiguous way between these two possible scenarios would have been
to obtain milliarcsecond resolution deep images of CSL-1. Such
image, collected by the authors on January 11 2006 using the ACS/WFC
on HST, showed beyond any doubt that CSL-1 is a pair of two
interacting galaxies (see the detailed discussion presented below and in
\citealt{saz06}).
This conclusion was confirmed
by an independent group of observers \citep{ago06}. In what follows
we present the results of the models which were implemented to
study the properties of CSL-1 and which appear to be of general
interest for future searches of cosmic strings.

The paper is organized as follows. In Section \ref{gravlens} we
give a short review of lensing by cosmic strings, emphasizing the
physical meaning of the phenomena. In Section \ref{images} we
discuss the morphologies obtained from detailed numerical
simulations, while Section \ref{CSL-1} is devoted to a detailed
discussion of the CSL-1 case based on the already mentioned HST
observations. Finally, in Section \ref{lenspair} we analyze the
chains of double images expected for the lensing by a cosmic
string.

\section{Cosmic string as a gravitational lens.}\label{gravlens}

As it was already mentioned, cosmic strings can be revealed by
means of gravitational lensing \citep{vil, string1} due to their
peculiar signatures, which are significantly different from those
expected for classical lenses. We wish to stress
that gravitational lensing appears to be crucial since it is the
only model independent observable quantities associated to cosmic
strings.

Photons from a background source move around the string and by
circum-navigating the string, they form two images on its sides.
Since along the two trajectories the space is flat, there is no
gravitational attraction exherted by the string on the photons and
no distortion is introduced. However, in spite of the fact that
the metric is locally flat, the global properties of the
space-time are not Minkowskian but conical, and a complete turn
around the position of the string, gives the total angle $\phi$
smaller than 2$\pi$, while the difference 2 $\pi - \phi$ is the
so-called "deficit angle $\Delta \theta$" defining the lensing
properties of the string. The physical properties of a cosmic
string predicted by Kibble are characterized by just one
parameter, namely the mass per unit length $\mu$, from which the
deficit angle $\Delta\theta=8\pi G \mu$ and the lensing properties
can be derived \citep{kib76, vil, string1, string2, shl05}.
In gravitational lensing processes the angular distance between
lensed images depends on the deficit angle and from the linear distances
(from the observer to the lens and from the observer to the background source).
In general this parameter also depends on the transverse velocity and
orientation of the string with respect to the observer; however in the 
simplified model derived here both of them can be safely neglected.

\subsection{The case of a point-like source}
In order to understand the main physics of the phenomenon, we
start from the simplest case: that of gravitational lensing by a
straight string,  to the line of sight and with zero
velocity. More complex properties of the string, such as its
velocity, curvature, possible charge, gravitational waves, etc.
can be found in literature \citep{string1, string4, dam04, shl05}
and will be treated in more details in forthcoming papers. For instance,
the hypothesis of a straight string fits well the case of CSL-1,
since this object shows circular and undistorted isophotes which
could not be explained in terms of a locally curved string.

The geometry of the phenomenon has been described in \cite{sch92,
saz98}, and will be shortly summarized here.

\begin{figure*}
\begin{center}
\includegraphics[width=10.0cm,angle=270]{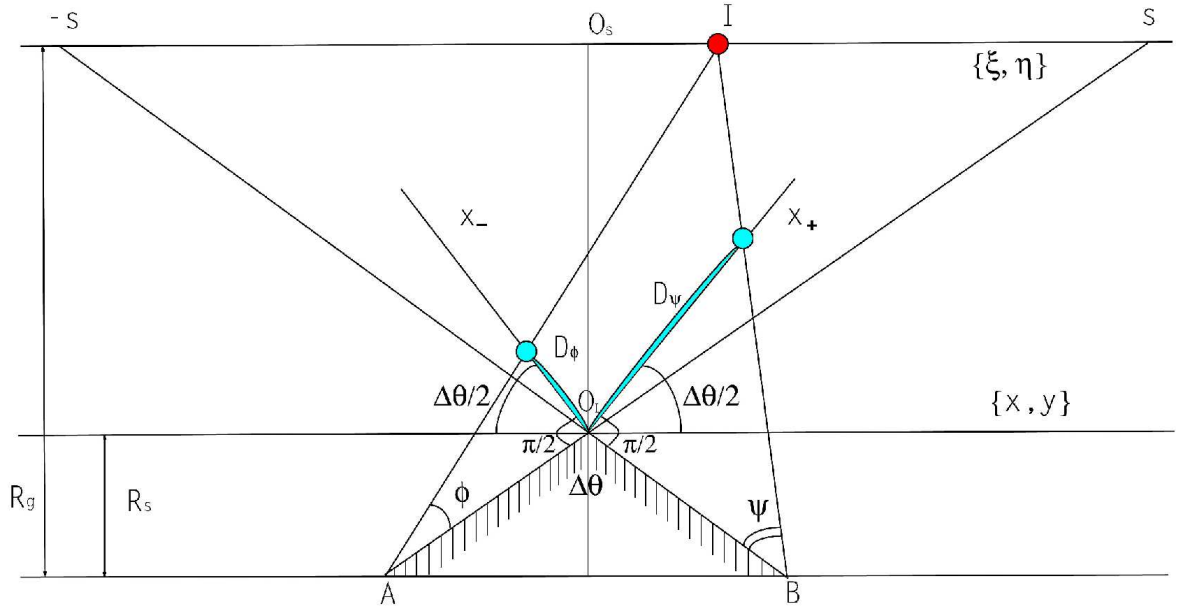}
\end{center}
\caption{The simplest geometric interpretation of gravitational
lensing if a background point-like object (I) is lensed by a cosmic
string. The dashed region marks the cut associated to a conical
space time once it it is seen in an euclidean space, and the points
A and B mark the apparent positions of the observer in the euclidean
space. In other words, the observer shall see two images (blue
circles) separated by the distance $D = D_{\phi}+D_{\psi}$. For the
other symbols, see the text.} \label{fig1}
\end{figure*}

In usual gravitational lens theory the main axis coincides with
the line joining the observer and the barycenter of the lens. In
our case the lens is a one dimensional object, and therefore we
may define (see Fig. \ref{fig1}) it as the shortest line which
connects the observer and the string. Let now us extend this line
to a background object and introduce three planes perpendicular to
such main axis The first one is the "object plane" which
intersects the center of the background source; the second one is
the "lens plane" which contains the nearest point of the
string, and, finally, the last one, the  "observer plane", which
contains the observer.

Let the background source be point-like. With reference to
Fig. \ref{fig1}, axes $O_S\xi , O_S\eta$ define the coordinate
system on the plane of the background source and the origin of this
coincides with the intersection of this plane with main axis.
The vector $\{\xi , \eta \}$ defines the distance from the
origin of object coordinate system to the position of the source
$(I)$. The axis $O_S\eta$ is perpendicular to the plane of Fig. \ref{fig1}.

On the lens plane we introduce the definition of axes with latin
characters. $O_Lx , O_Ly$ define the coordinate system in the plane of
the string (again, $O_Ly$ is perpendicular to the plane of
Fig. \ref{fig1}) and $x_{-}$ and $x_{+}$ denote respectively the
left and right parts of $O_Lx$ axis the string plane and coincide
with the axis $O_Lx$  when the points A and B are brought
together; $\Delta\theta$ is the already introduced deficit angle,
$R_s$ is the distance between the observer and the string plane,
and $R_g$ is the distance between the observer and the source. In
this geometry and under our assumptions, the observer will see
the double images of a background source separated by the angular 
distance $\delta \theta$:
\begin{eqnarray*}
\delta \theta = \Delta\theta \dfrac{R_g-R_s}{R_g}
\end{eqnarray*}

\begin{figure*}
\begin{center}
\includegraphics[width=12.0cm]{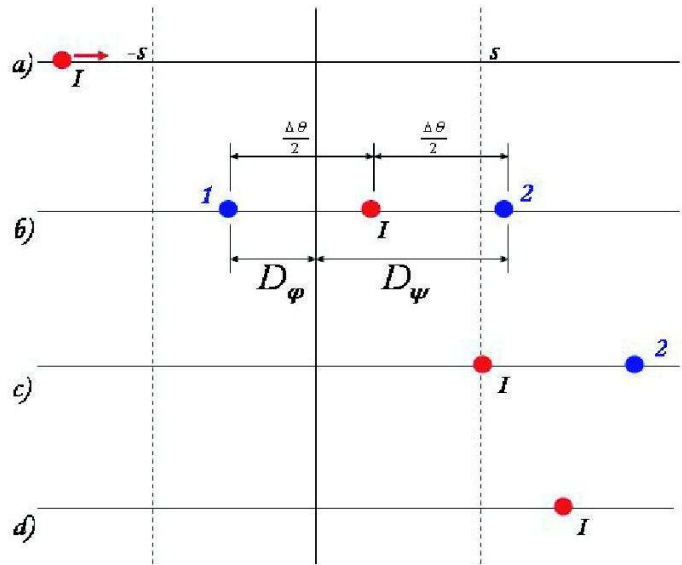}
\end{center}
\caption{The aspect of the images of a background source as a
function of the relative position of the source and of the string
strip $[-s,s]$ (done on string plane).} \label{fig2}
\end{figure*}

\noindent Depending on the position of the background source
(Fig. \ref{fig2}) the observer will see one or two images. If the
background source $(I)$ falls inside the strip $[-s,s]$, the
observer will see two images on the string plane (we wish to
stress that, in the euclidean space, this corresponds to the fact
that the observer consists of two points A and B).

The lensing equation relates the physical distances (positions on
the lens plane) $D_\phi$, $D_\psi$ with $\xi_s , \eta_s$
(positions on the source plane), as function of the deficit angle
$\Delta\theta$, $R_g$ and $R_s$. Being the deficit angle very
small, it is possible to derive a simple relation between angles:
\begin{eqnarray*}
\phi+ \psi\approx \Delta\theta \biggl( 1-\frac{R_s}{R_g} \biggr)
\end{eqnarray*}

\noindent The angles $\phi$ and $\psi$ are defined as:

\begin{eqnarray*}
\psi = \frac{1}{2} \biggl( \frac{2\xi_s}{R_g}+\Delta\theta
\frac{R_g
- R_s}{R_g}  \biggr) \\
\phi = \frac{1}{2} \biggl( -\frac{2\xi_s}{R_g}+\Delta\theta
\frac{R_g - R_s}{R_g} \biggr) \label{angls}
\end{eqnarray*}

\noindent and:

\begin{eqnarray*}
s = \Delta\theta (R_g - R_s)
\end{eqnarray*}

\noindent If we omit the second order term, the physical distances
$D_\phi$, $D_\psi$ can be written as:
\begin{eqnarray*}
D_\psi = R_s \psi \mbox{\hspace{2cm}} D_\phi = R_s \phi
\label{sgl}
\end{eqnarray*}

\noindent The lens equation can then be derived from the following
equations:
\begin{equation}
x_1 = \dfrac{R_s}{R_g} \bigl(\xi_s+ \frac{s}{2}\bigr)
\mbox{\hspace{2cm}} y_1 = \dfrac{R_s}{R_g} \eta_s \label{lenseq1}
\end{equation}

\begin{equation}
x_2 =  \dfrac{R_s}{R_g} \bigl(\xi_s - \frac{s}{2}\bigr)
\mbox{\hspace{2cm}} y_2 = \dfrac{R_s}{R_g}\eta_s \label{lenseq2}
\end{equation}

\noindent where $x_1 , y_1$ and $x_2 , y_2$ are the coordinates on
the lens plane of the first and second images, respectively.

Therefore, in the case of a point source falling inside the string
strip, the observer will observe two identical images of the
source, with positions defined by the string lens equations
(\ref{lenseq1} - \ref{lenseq2}) and, as long as the source is
point--like and the photon beams move in a quasi Euclidean space,
the two images will have identical optical properties.

\subsection{The case of an extended source}

The width of a cosmic string strip, defined by its deficit angle,
depends on the string linear density (or tension) $\mu$. However the width
of the cosmic string itself (or its cross section) is negligible small
($10^{-17} \div 10^{-33}$ cm) being compared with the size of any
astronomical object, because its mass scale is not less then 1
TeV.

Thus the size of any extragalactic source is much larger than the
width of the string, and any source can be regarded as extended in
comparison with the string size. In this case, the general
equation of mapping by a string is given by $I(x, y)=$:
\begin{eqnarray*}
I(\dfrac{R_s}{R_g} \bigl(\xi_s + s/2\bigr), \dfrac{R_s}{R_g}\eta_s) & & s<\xi_s\\
I(\dfrac{R_s}{R_g} \bigl(\xi_s + s/2\bigr), \dfrac{R_s}{R_g}\eta_s) + I(\dfrac{R_s}{R_g} \bigl(\xi_s -s/2\bigr), \dfrac{R_s}{R_g}\eta_s) &  \mbox{for} & -s \le \xi_s \le s\\
I(\dfrac{R_s}{R_g} \bigl(\xi_s - s/2\bigr), \dfrac{R_s}{R_g}\eta_s) &  & \xi_s < -s\\
\end{eqnarray*}

\begin{figure*}
\centerline{\includegraphics[width=8.0cm]{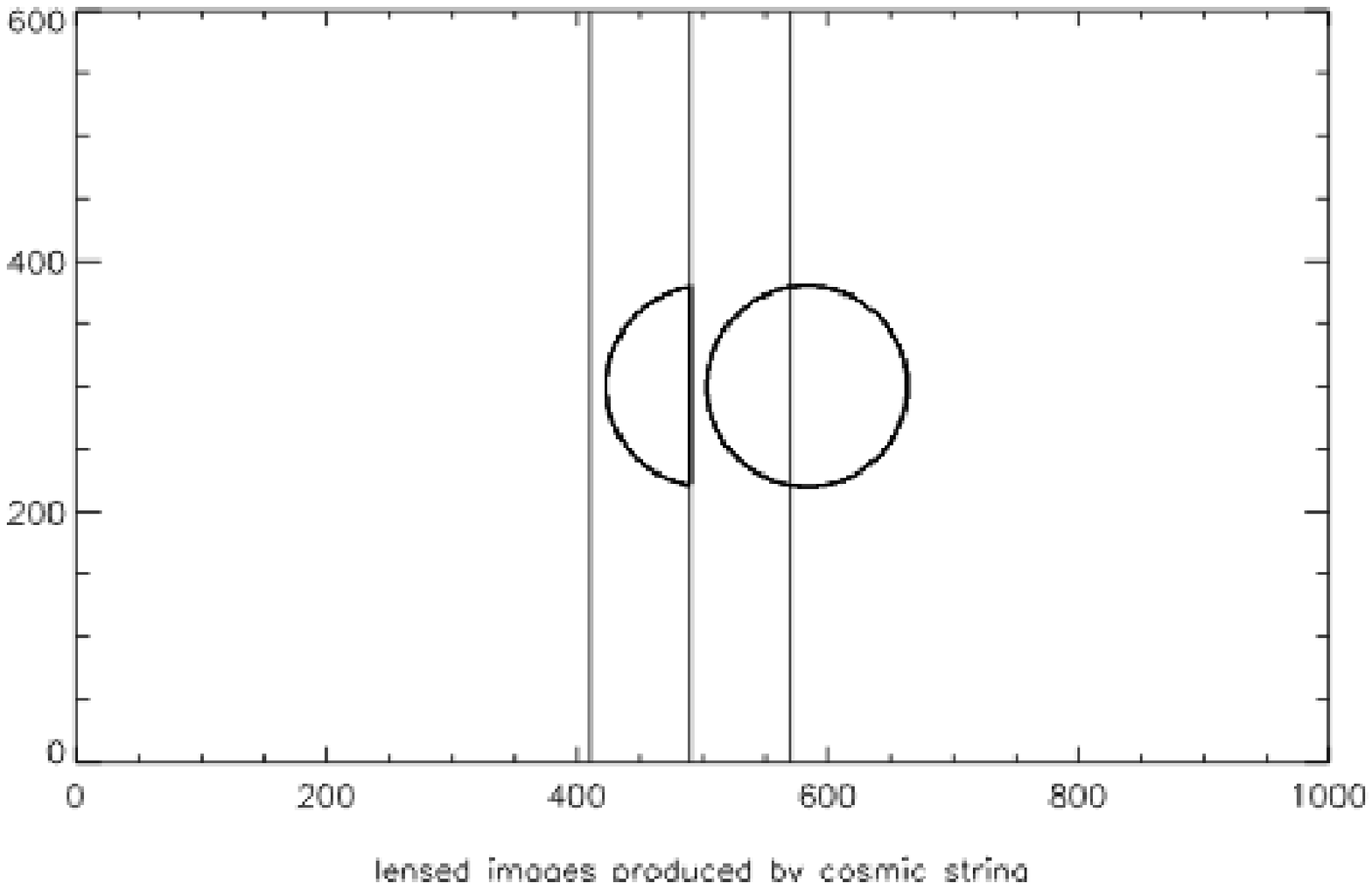}
\includegraphics[width=8.0cm]{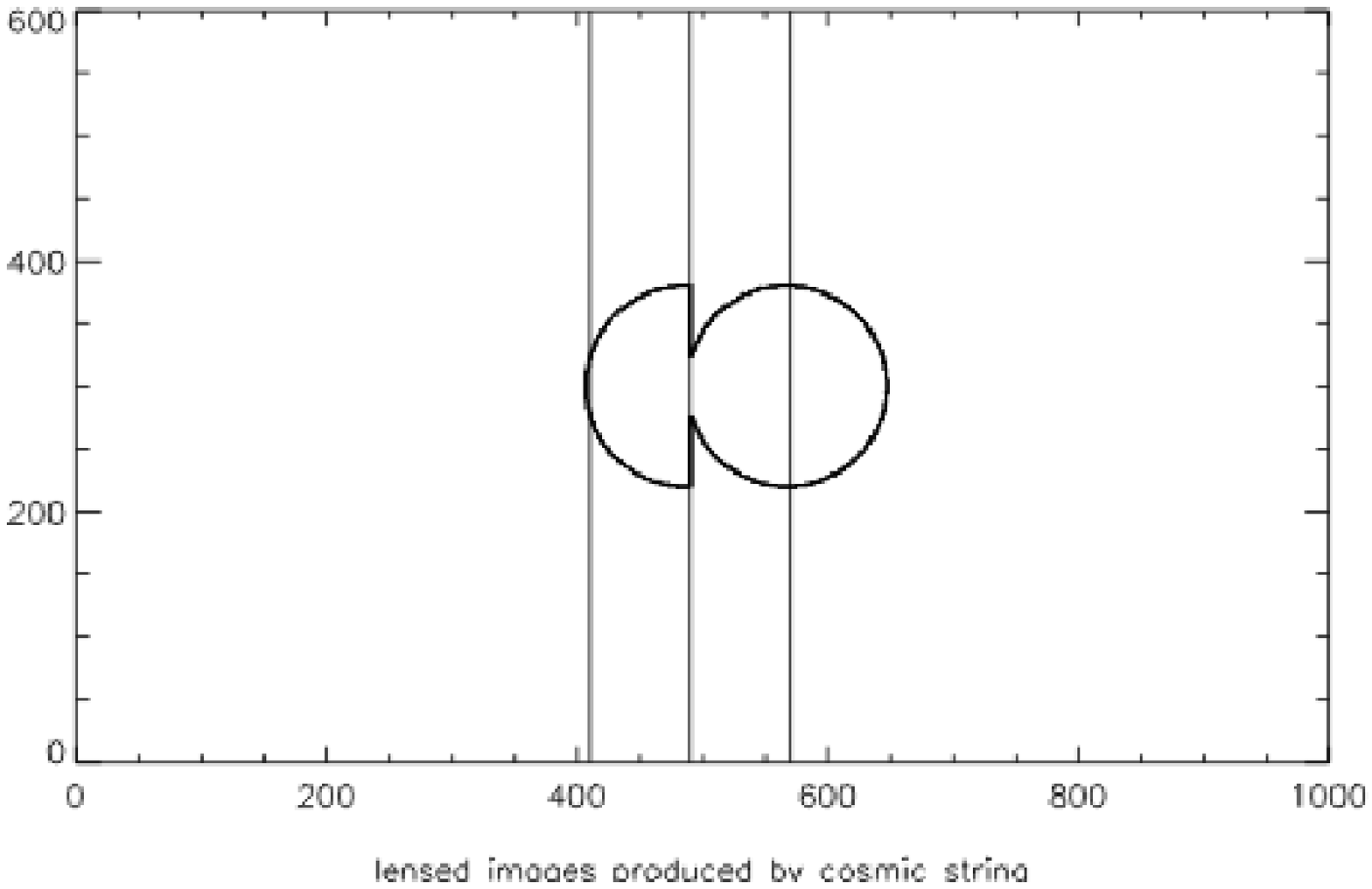}}
\caption{Left: the images of a circular source lensed by a cosmic string. The straight solid line represent the position of the
string, while the dashed lines show the position of string strips relative to string itself. Right: the images of a circular source
lensed by a cosmic string. The position of string and Einstein strips are the same as in previous figure. The only difference is
that the radius of the source is larger than the angular distance between source center and string. } \label{fig3}
\end{figure*}

\noindent For each point of the source we can follow the same
procedure described in the previous paragraph and, if the point is
inside the string strip, it will be displayed on the other side of
the string, while, if it is not, it will be cut away thus
producing sharp edges in the isophotes of the source images.
Fig. \ref{fig3} shows an example of what would happen to a
circular source lensed by a string. Notice that the sharp
edge introduced by the string is clearly visible.
In order to better quantify such effect, let us assume an
homogeneous brightness distribution over the disk of the source.
It is worth to stress that this assumption does not affect much
the generality of the results, since in the case of a source with
a radial dependence of the brightness distribution, the source can
be approximated as a combination of rings of different brightness
and the result can be obtained by integrating over the rings.

Let now the axis of the coordinate system be oriented along the
string ($Oy$) and perpendicularly ($Ox$) to it  ($O_L \equiv O$).
The source coordinates will map onto the lens plane in the same
way: $\xi$ will coincide with $Ox$, and $\eta$ with $Oy$. It is
useful to note explicitly that the origin of the source coordinate
system will map into the origin of the lens coordinate system
$(\xi=0 , \eta=0) \ \rightarrow \ (x=0, y=0)$.

Suppose also that the source has circular shape with radius
$\rho_s$ and center in $(\xi_s , \eta_s)$. The outer contour is then
described by the equation:
\begin{equation}
(\xi -\xi_s)^2 + (\eta -\eta_s)^2 = \rho_s^2 \label{sh1}
\end{equation}

\noindent and the center of the circle will map into:
\begin{eqnarray}
x_{1s} = \dfrac{R_s}{R_g} \bigl(\xi_s+ \frac{s}{2}\bigr)
\mbox{\hspace{2cm}} y_{1s} = \dfrac{R_s}{R_g} \eta_s \\
x_{2s} =  \dfrac{R_s}{R_g} \bigl(\xi_s - \frac{s}{2}\bigr)
\mbox{\hspace{2cm}} y_{2s} = \dfrac{R_s}{R_g}\eta_s
\label{lenseq3}
\end{eqnarray}

\noindent where $1$ and $2$ refer, respectively, to the first and
second image. The radius in the lens plane becomes:

\noindent $ r_i =\dfrac{R_s}{R_g}\rho_s $

\noindent and the outer boundary is described by the equations
\begin{equation}
(x -\dfrac{R_s}{R_g} \bigl(\xi_s \pm \frac{s}{2}\bigr))^2 + (\eta
- \dfrac{R_s}{R_g} \eta_s)^2 = r_i^2 \label{shl}
\end{equation}

\noindent where the sign differentiates between the first ($+$)
and the second ($-$) image.

In fact, an observer does not know the true position of the source
in the sky. It can be reconstructed in most cases, but in some
cases the reconstruction is not unique. In the simple case when
the radius of a source is less then the angular distance between
the source center and the string we will define as first image the complete one,
while the second will be the incomplete one, as one can see in
Fig.~\ref{fig3}.

The situation becomes more difficult if the radius of the source
(or radius of a ring of the source) becomes larger than the
distance between source center and the string.

\noindent If part of the first image intersects the string
position, all points at $x<0$ (if they obey the eq.
(\ref{shl})) need to be cut away and the corresponding part of
circle turns into a straight line coinciding with the string
position.

\noindent The same is
also true for the second image, but inverted: the visible part
being that for which $x<0$. In other words, all points obeying
eq.~(\ref{shl}) and for which $x \le 0$, need to be cut out and
replaced with a straight line coinciding with the string position.
The edge in the first image appears if the radius of the circle is
larger than $\rho_s > \eta_s +s$ (see Fig.~\ref{fig3}, right panel). We shall
therefore assume $\eta_s > 0$.

The linear size of the edge can be written as:

\begin{eqnarray*}
\Delta y_1 = 2\dfrac{R_s}{R_g} \sqrt{\rho_s^2 - \bigl( \eta_s +s
\bigr)^2} \label{cut1}
\end{eqnarray*}

\noindent When the edge is absent, the total flux from the source
is proportional to the source area $\sim \pi \rho_s^2$.

In the opposite case, the total area is smaller and becomes:
\begin{eqnarray*}
A_1 = \bigl( \pi - \phi_1 + \frac{1}{2}\sin 2\phi_1\bigr) \rho_s^2
\label{area1}
\end{eqnarray*}

\noindent where:

\noindent $ \sin \phi_1 = \dfrac{\Delta y_1}{2r_s}.$

\noindent Also for the second image the edge is defined by the
condition $x=0$, and the size of the edge is given by:
\begin{eqnarray*}
\Delta y_2 = 2\dfrac{R_s}{R_g} \sqrt{\rho_s^2 - \bigl( \eta_s -s
\bigr)^2} \label{cut2}
\end{eqnarray*}

The condition $\rho_s \le |\eta_s -s|$ must then be matched in
order to produce the edge in the second image.
This inequality is not uniquely defined.
In fact, if the center of the source falls outside the Einstein
strip ($\eta_s > s$), the center of the second image is to the
right hand side of the string, and an observer sees less then half
of the circle (case A).
If the center of the source is inside the Einstein strip ($\eta_s
< s$), then the center of the second image is on the left hand
side of the string, and an observer sees more than half of the
circle (case B).
In both cases the sizes of the edges are equal.

\noindent In case A, the visible area is equal to:
\begin{eqnarray*}
A_{1A} = \bigl( \phi_2 - \frac{1}{2}\sin 2\phi_2\bigr) \rho_s^2
\label{area2A}
\end{eqnarray*}

\noindent where

\noindent $ \sin \phi_2 = \dfrac{\Delta y_2}{2r_s}.$

\noindent Instead, in case B, the visible area is equal to:
\begin{eqnarray*}
A_{2B} = \bigl( \pi -\phi_2 + \frac{1}{2}\sin 2\phi_2\bigr)
\rho_s^2 \label{area2B}
\end{eqnarray*}

If the first image does not produces an edge, while the second one
does (see for instance Fig.~\ref{fig3}), the size of the edge will
be equal to that in the second image. If, instead, the edge in the
first image does exist, the total size of the edge will be equal to
the difference between the edges in the first and second image
(see Fig.~\ref{fig3}, right panel). This remark is crucial to understand ground based
observations. In fact, in this case we need to probe very low surface
brightness isophotes in order to detect the edge.

Furthermore, since these isophotes will usually have large radii,
the edges in the two images will merge and the resulting appearance
will be given by the difference between the two edges.

\begin{figure*}
\includegraphics[width=8cm]{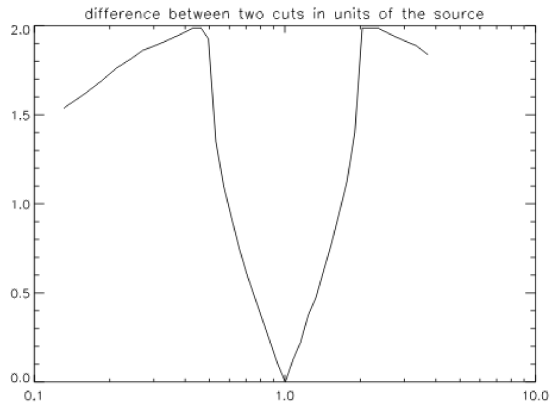}
\caption{ Graph showing the difference between two edges (edge of
first and second images) as function of images intensities. The
difference value is plotted along vertical axes. The ratio of
intensities of two images is plotted along the horizontal axes and the scale
is logarithmic} \label{fig5}
\end{figure*}

Fig.~\ref{fig5} shows the difference between the two edges as a
function of the intensity ratio of the two images ($F$). One can
see that $F=1$ corresponds to zero difference.
Generally speaking the case where the value of $F$
is around unity is very hard to disentangle (especially in presence of noise)
from that of a chance alignment of two similar
looking galaxies.

Fig.~\ref{fig7} presents the effects produced by a typical string 
(whith mass scale of the order of $10^{15}$ GeV) upon a background galaxy at
redshift z$\sim$ 0.5, producing splitted images $\sim 2$ arcsec apart.

If the ratio falls within the $0.9 < F < 1.1$ range, the
difference between the edges is smaller than 0.1 of the source
radius ($\leq$ 0.2 arcsec in our case) and therefore very high angular
resolution is required in order to detect it.
The difference increases when the ratio $F$ increases (or
decreases) and, for instance, when it is $\sim 2$, the difference
is almost equal to the radius size.
The above discussion confirms that the detection of sharp edges of
pairs of lensed images along the position of the string is, at
least in theory, possible also from the ground.

We wish also to stress that one of the most characteristic features of lensing by a cosmic
string is the fact that all details (such as galactic arms, bright
spots, globular clusters, supernovae, etc.) which are present in
the first image, will also be reproduced in the second one if they
fall inside the string strip.

An additional feature appears if we take into account the possible
time delay between two images, which is determined by the difference between
the two photon paths (AI and BI, see Fig.~\ref{fig1}).

\begin{eqnarray*}
BI=\frac{R_g}{\cos{(\Delta\theta/2-\psi)}},
\end{eqnarray*}

\begin{eqnarray*}
AI=\frac{R_g}{\cos{(\Delta\theta/2-\phi)}}.
\end{eqnarray*}

\noindent and the difference between the two paths can be written
as:

\begin{equation}
\Delta L = \dfrac{1}{2}R_s \Delta\theta (\psi -\phi)
\label{del}
\end{equation}

\noindent where $\Delta L=AI - BI$. This difference can also be
written in terms of the coordinates in the source plane:

\begin{eqnarray*}
\Delta L= \dfrac{R_s}{R_g} \xi_s \Delta\theta
\end{eqnarray*}

The best way to represent this value is in observable terms. In
eq.~(\ref{del}) only one term $R_s$ has to be expressed in terms of
$\sim H^{-1}$ to get the time delay expressed in observable
values:

\begin{eqnarray*}
\Delta t = \dfrac{1}{2H} f(z_s, \Omega_m, \Omega_{\Lambda})
\Delta\theta (\psi -\phi) \label{delay}
\end{eqnarray*}

\noindent where $H$ is the Hubble parameter, $z_s$ is the redshift
of string, $\Omega_m, \Omega_{\Lambda}$ are the contributions of
matter and dark energy respectively, and $f$ is a function which
describes the cosmic distance to the string.


When dealing with time delays, a possible source of
misinterpretation could be the presence of a variable object
within the source. In the case of a supernova, for instance, the
time delay between the two images would become important since,
should it be  greater than the characteristic variability time of
supernovae, it could be seen in one image and not in the other.

\section{Simulated images produced by a cosmic string.}\label{images}

In order to produce realistic simulations of the effect described
above, we made use of a "virtual" galaxy obtained using a de
Vaucouleurs surface brightness profile \citep{GV_53}:
\begin{eqnarray*}
I(r) =I_v exp(-7.6692(\dfrac{r}{r_{ch}})^{1/4})
\end{eqnarray*}
\noindent truncated at $r > 10 r_{e}$ in order to speed up
computations. To be as realistic as possible, we used the
redshift, apparent magnitude in the Johnson $V$ band and effective
radius derived for CSL-1 in \cite{saz05} which are equal to
$z=0.46$, $V=21.05$ and $r_e = 1.6"$, respectively. As
observational parameters we assumed those adopted in our HST
observations of CSL-1 (which are rather typical). We assumed a
pixel size of 25 mas, i.e the pixel size achievable with HST and
typical dithering, and convolved the model with a FWHM=0.1" PSF to
simulate the angular resolution expected in the F814 band (which
roughly corresponds to the rest-frame V band). We used a
stochastic process to compute the Poissonian noise per pixel,
using the expresion:

\begin{equation}
\sqrt{ (C+B_{sky}+B_{det})t+n_{read}N^2 }
\label{noise}
\end{equation}

\noindent where $t=14$ ks is the total exposure time,
$C$ is the signal from the astronomical source in
counts/second, $B_{sky}$ and $B_{det}$ are the average sky and
detector background, $N$ is the readout noise and $n_{read}$ is
the number of CCD readouts. The actual values of $B_{sky}$,
$B_{det}$ and $N$ were obtained from the ACS Instrument
Handbook\footnote{http://www.stsci.edu/hst/acs/documents/handbooks/cycle15/cover.html}.

\begin{figure*}
\centerline{\includegraphics[width=14cm]{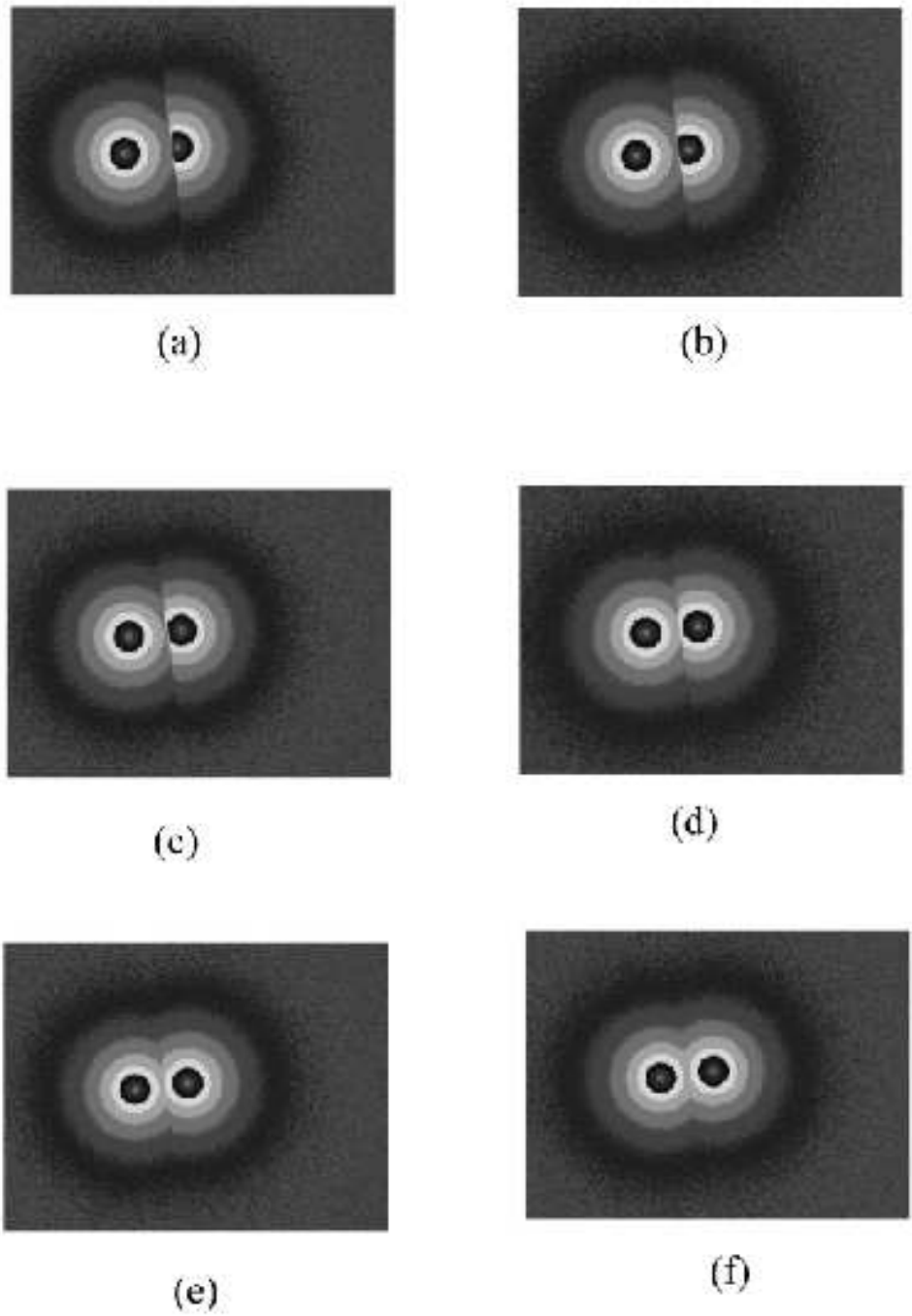}}
\caption{The images of a source lensed by a cosmic string. We assume
de Vaucouleurs profile brightness distribution over disk.
Each picture represent one step relative to position
of the string. Noise is included (see text).} \label{fig7}
\end{figure*}

Note that when multiple observations are dithered and stacked the
actual noise statistics is not simply represented by the
expr. (\ref{noise}) due to correlation among pixels on scales
given by the dithering pattern \citep{casertano00,fruch02}.
However for our observations this results in a noise suppression
factor of $\sim 2$ which can be compensated by rebinning as long
as the lensing signatures are large compared to the pixel scale.
Furthermore the comparison of Fig. \ref{fig7} and \ref{CSL1_HST}
shows that expr. (\ref{noise}) is adequate for the simple
model discussed here.

The simulated elliptical was then placed within the lensing strip at different
angular distances with respect to the string.
In Fig. \ref{fig7} we show the results of our simulations.
The figure is composed by 6 panels (a through f) corresponding to
intensity ratios $F$ equal to $1.4$, $1.27$, $1.18$, $1.10$,
$1.04$, and $0.99$, respectively. The latter value corresponds to
an almost symmetric situation, in which the observer will hardly see
the sharp edges produced by the string even with high (HST like)
angular resolution.

We notice that in the case of CSL-1 the intensity ratio of the two
components falls in the range $1.06-1.04$ and therefore roughly
corresponds to panel (e). In the images,
the sharp "edges" introduced in the outer isophotes by the string
are apparent.

\begin{figure*}
\centerline{\includegraphics[width=10cm]{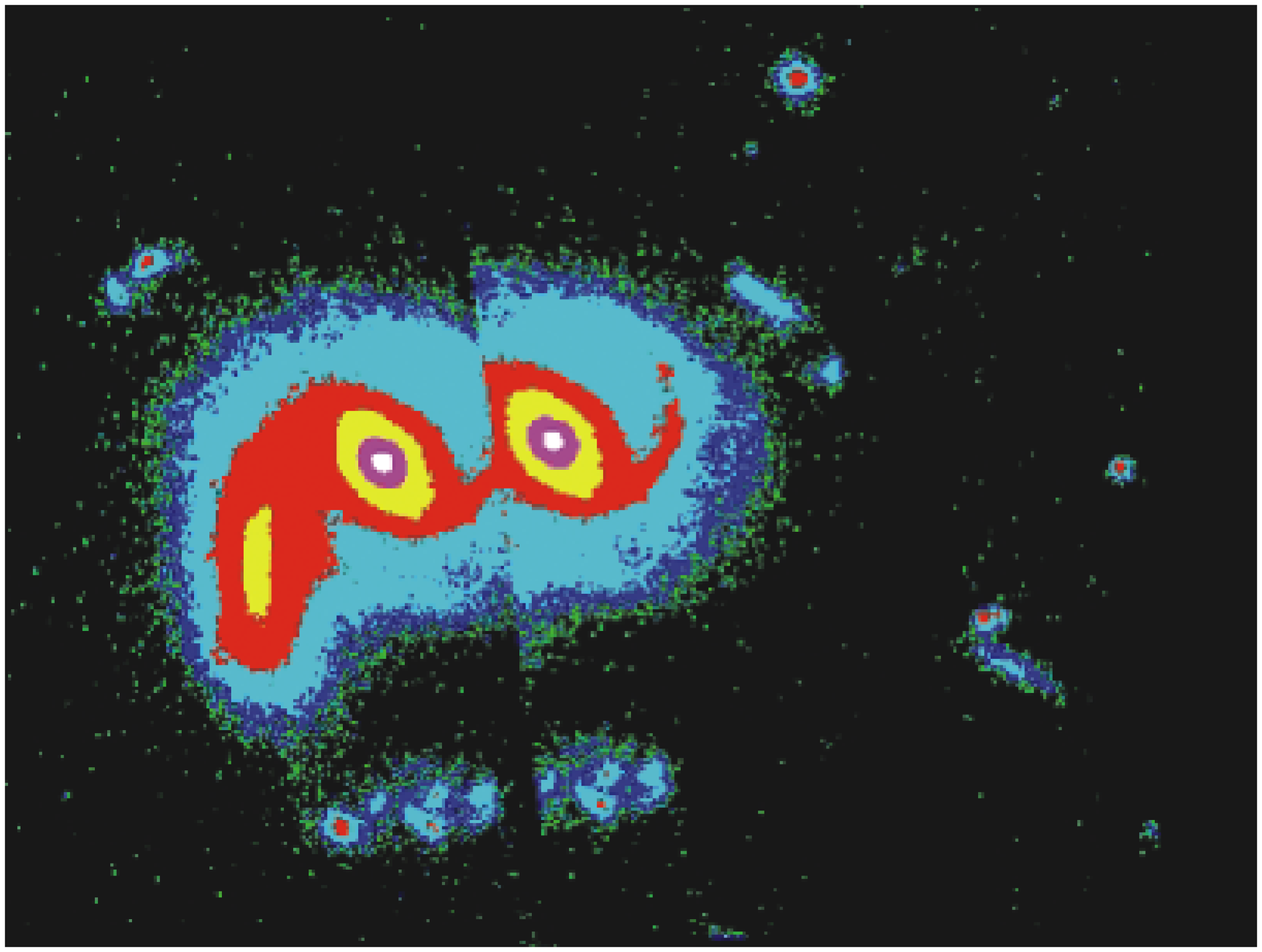} a)}
\centerline{\includegraphics[width=10cm]{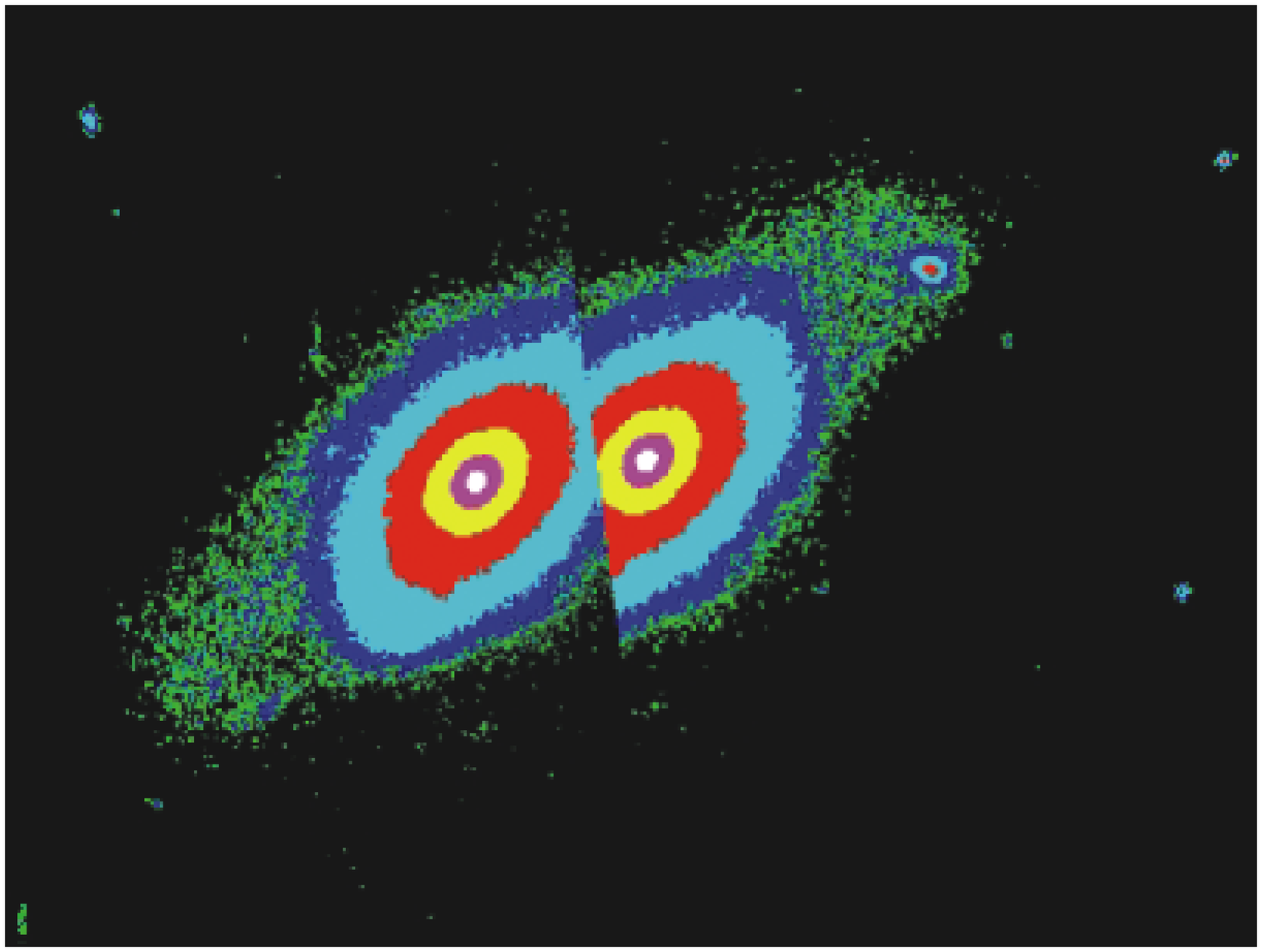} b)}
\centerline{\includegraphics[width=10cm]{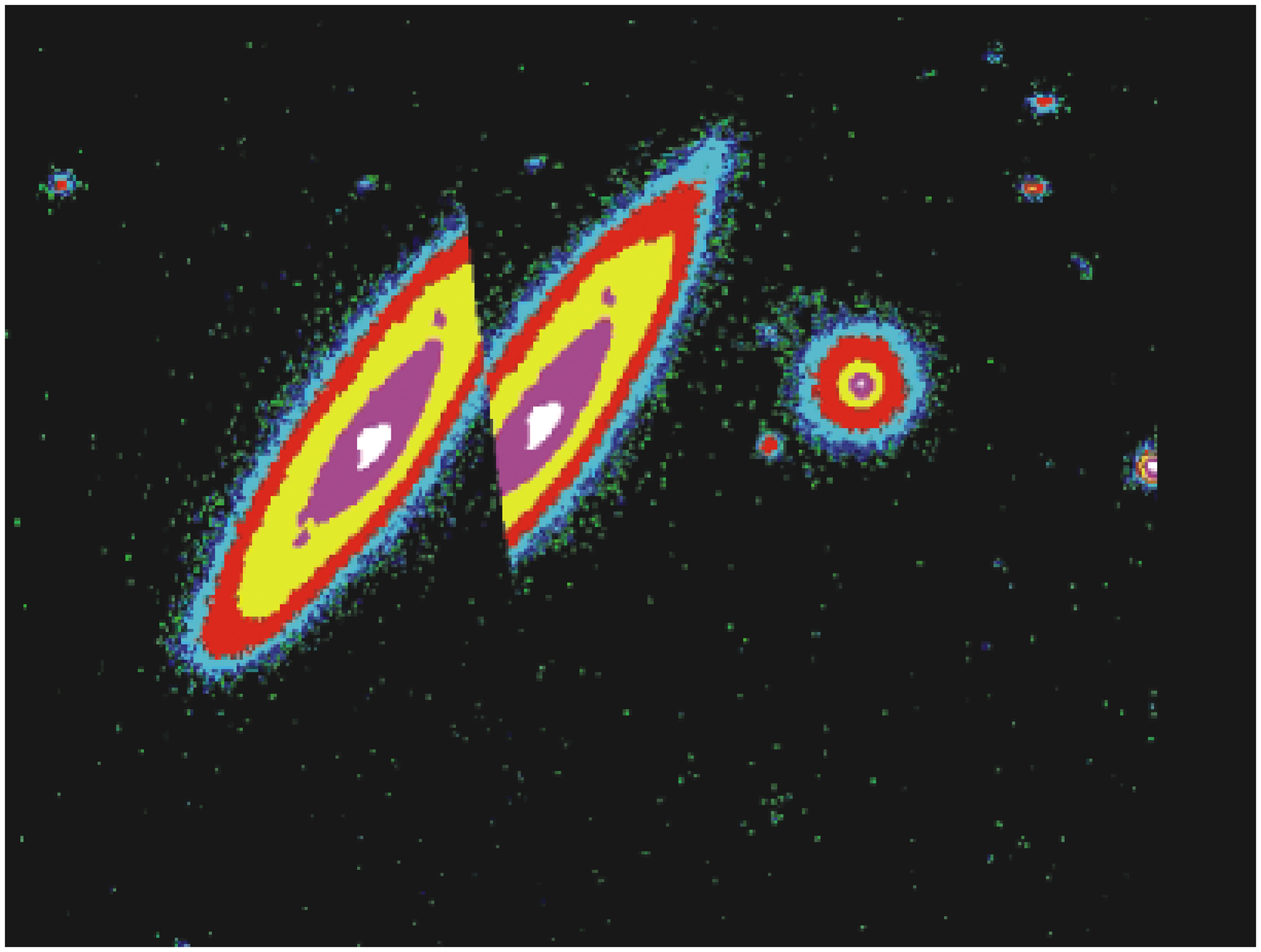} c) }
\caption{Panel a: this picture represent the lensed image of a spiral galaxy. The direction of a string is almost perpendicular
to galaxy plane. Duplicated details are clearly visible. Panel (b): in this case, the string is inclined with respect to the galaxy
plane. As a result sharp edge appears. Panel (c): Also in thei case the direction of string is inclined with respect to galactic
plane.} \label{fig8}
\end{figure*}

For completeness, we also present the lensed images of a set of
 three spiral galaxies extracted from our HST data (Fig. \ref{fig8}).

\section{HST image of CSL-1.}\label{CSL-1}
\begin{figure*}
\includegraphics[width=10cm]{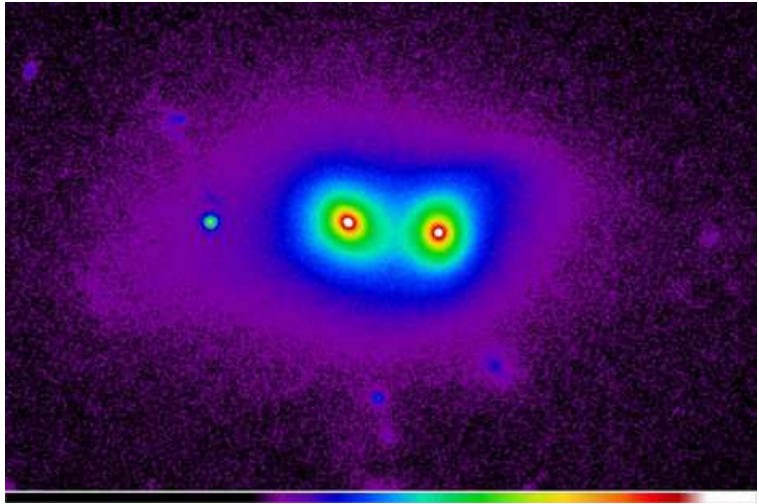}
\caption{CSL-1 image in pseudocolor as obtained by stacking (see
text) 6 HST orbits obtained on January 11, 2006 in the F814W band.
} \label{CSL1_HST}
\end{figure*}

To test whether CSL-1 was actually a lens produced by a cosmic
string we observed the double source with the HST/ACS camera
during Cycle 14, using Director's Discretionary Time. CSL-1 was
observed for 6 HST orbit in the F814W band (comparable to
Johnson-Cousins I-band) yielding an effective exposure time of
$\sim 14000$ seconds. The observations were performed adopting a
1/3 pixel dither pattern, to allow sub-pixel sampling of the HST
PSF and accurate cosmic ray rejection. All 6 orbits were combined
through the Multidrizzle software \cite{koe02} using a 1/2 pixel
(0.025 arcsec/pixel) resampling pattern. In Fig.~\ref{CSL1_HST} we
show the final stacked image.

As it can be seen by comparison with our simulations (panel (e) of
Fig.~\ref{fig7}) in the HST data there is no sign of the peculiar
features (sharp edges) predicted in the case of lensing by a
cosmic string. The faint isophotes of the two components have
different shapes, which is incompatible with CSL-1 being lensed by a cosmic
string. In fact in the cosmic string scenario all morphological
features of the source falling inside the deficit angle, would be
mirrored on the opposite side of the string. However in the HST
image we do not see such mirroring effect for the two components,
nor for any other faint feature which, would have fallen inside
the deficit angle of the string and should have been duplicated,
e.g. the faint sources on the southern side of CSL-1 visible in
the right panel of Fig. \ref{CSL1_HST}.

\begin{figure*}
\includegraphics[width=10.0cm]{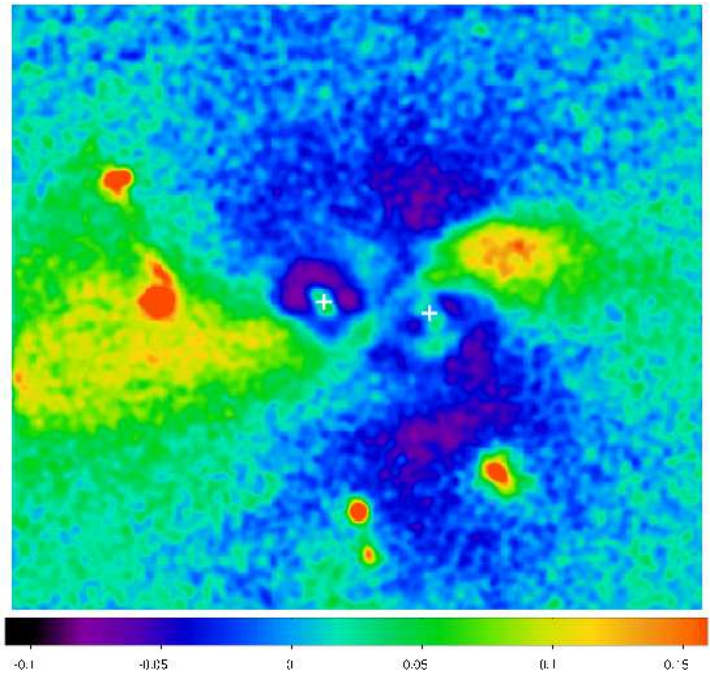}
\caption{The normalized residuals (residuals/model) obtained by
subtracting from the HST images a model consisting of two de
Vaucouleurs light profiles. Crosses mark the centroids of the two galaxies.} \label{res}
\end{figure*}

To further check whether the distortions observed in the faint isophotes
are caused by tidal interactions between the two ellipticals we
fit the two objects with two de Vaucouleurs $r^{1/4}$ light
profiles and subtract the model from the original data. The
residual image, presented in Fig.\ref{res}, clearly shows the
presence of warped structures in the CSL-1 outskirts, most
probably tidal tails due to the interaction between the two
galaxies. The detailed photometry of the objects will be discussed
elsewhere (Paolillo et al. in preparation).

\section{How many lensing pairs we have to expect?}\label{lenspair}
As discussed in Sec.\ref{introduction}, the most evident signature
of a cosmic string is to produce a strip of multiple images along
its path. This would be the first feature to look for in any
dedicated search for cosmic strings within large astronomical
surveys. As template cases, in what follows we derive the expected
number of lensed images using as template case the CSL-1 field as
it appears in the $R$ band mosaic taken from the OAC - Deep Field
(OACDF) \citep{cap1, alc04} and in the deeper observations
obtained with HST.

The presence of a background galaxy inside the deficit angle of a
string is a stochastic process determined by the area of the
lensing strip and by the density (number of objects per unit solid
angle) of background galaxies. The larger is the field of search,
the larger is the number of lensed objects that should be found.

All lensed objects will fall inside the narrow strip defined by
the path of the string and by the deficit angle. Along this path,
an observer should therefore see a sort of "milky way" of double
images of galaxies. Historically speaking, this effect was first discussed
by \cite{vil,vil84,vil86,string2,string5}, and we shall just summarize it
briefly in the framework of a simple model. For the sake of
simplicity, we shall consider all background object as point--like
sources. In the case of a straight string, one can easily estimate
the expected number of lensed galaxies as

\begin{equation}
\left< N \right> = n_g 2 l \Delta \theta  \label{eq1}
\end{equation}

\noindent Here $n_g$ is the density of galaxies per unit solid
angle, $\Delta \theta$ is the deficit angle of the string, and $l$
is the length of the string in the chosen field. Both $\Delta
\theta$ and $l$ are expressed as angular measures. A more complex
case emerges if the string is assumed to be curved \cite{string5}.
A simple estimate can be derived as it follows. The lenght of a
curved string is larger than that of a straight one; therefore,
the lensing strip will cover a larger area on the sky in the same
patch and its lenght can be written as \citep{string5}: \[ l = \varrho
\left(\frac{\varrho}{l_{c}}\right)^a \]

\noindent Here $\varrho = |\vec r - \vec r_1|$ is  the distance from the
point $\vec r$ to the point $\vec r_1$. $l_c$ is the correlation
interval. The parameter $a$ varies between $0$ (straight string)
and $1$ (in the case of random walk of the string); the last value
corresponding to purely brownian motion ($\varrho \sim \sqrt{l}$).

In the case ($a=1$), the expected number of lenses is:
\[ \left< N \right> =  2 \frac{\Delta \theta}{l_{c}} n_g A \]

\noindent where the product of the angular area $A$ of the patch
times the surface density of galaxies $n_g$ gives the number of
galaxies expected in the patch. Therefore, in the case of a
straight string, the minimum number of lensed objects is
proportional to the number of galaxies falling inside the string
strip.

In order to to estimate such figures, and compare them with what
is actually observed in the CSL-1 field, we must derive the number
of galaxies brighter than the assumed limiting magnitude in the
$R$ and F814W band.

Counts in the $R$ band can be obtained from the
existing literature, such as the moderately deep data by
\cite{couR}. Deeper counts were obtained \citep[cf.][]{cou} at
slightly different wavelength, and they need therefore to be
interpolated. Additional information, for the F814W filter can be
found in \cite{hubbledf1}, \cite{gar98}, \cite{sha98},
\cite{gar00}, based on the Hubble Deep Field. Using
the \cite{couR} counts and extrapolating them to $m = 24$ in the
$R$ band, we derive that in the OAC-DF, in a field of 16\arcmin
$\times$16\arcmin, there should be $\sim 2200$ galaxies having
magnitudes in the range $20 < m_R <24$. Comfortably enough, this
figure matches the number of extended sources actually detected in
the OAC-DF.

Using the above estimate, in the case of a straight string we
expect at least 9 lenses, while in the case of a random walk
string, the expected number is much larger: $\sim 200$. Obviously,
in the same region of the sky, also lenses produced by galaxies or
conventional lenses should be present, and their average density
can be derived through the product of the optical depth due to
lensing, times the number of galaxies in the field \citep[][cf.]{fuk92,koc93,chi99,ofe03}.
These estimates lead to an expected number of $\sim 2$
conventional lenses within the same magnitude range as above and
within the same area.

In the case of HST observation the number of lensed pairs should
decrease due to the smaller field of view and increase due to the
fainter limiting magnitude ($\sim 28$ in the F814W band and for
point like sources). For limiting magnitude the signal to noise
ratio is equal to 9 roughly. In this case the number of galaxies
per unit of solid angle is (\citet{Williams96}): is $n \approx 10^6 deg^{-2}$
for magnitude $AB \le 28$.

The field of view of the ACS/WFC on HST is roughly 3.5\arcmin
$\times$ 3.5\arcmin, so that the maximum length, for a straight
strip crossing diagonally the FOV, is $\sim 5 \ arcmin$. Assuming
that the width of the string strip is $\sim 2 \ arcsec$ as we
already discussed above,  eq. (\ref{eq1}) gives an average number
of $\sim 40$ lensed pairs within the HST field. The HST image of
the CSL-1 field in \cite{saz06} shows no trace of an excess of
galaxy pairs, further ruling out the existence of a cosmic string
in the field.

\section{Conclusions}

In the present work, we presented a detailed analysis of the observable effects induced
by the gravitational field of a cosmic string and tested it against our recent HST observations
of the lens candidate CSL-1.

Our observations proved, beyond any doubts, that CSL-1 is a
rather peculiar pair of interacting ellipticals and its detailed
photometry will be presented elsewhere (Paolillo et al in
preparation). 
The results of our analysis lead to some
general conclusions which will be useful in future searches for
possible gravitational signatures of cosmic strings to be
performed in existing or future digital surveys.

It is likely \citep{all, J.Polchinski} there are a few dozen
long strings crossing horizon volume and therefore, any survey
aimed at detecting them through the photometric signature induced
by the gravitational lensing phenomenon needs to be multiband,
very deep and of high photometric accuracy. Our simulations showed
that, while high angular resolution (HST like) is not required to
produce lists of candidates, it is definitely needed in order to
disentangle whether these candidates actually are the signatures
of a string and to constrain the physical properties of the
string.

\section*{Acknowledgments}
The authors wish to thank the Director of the HST Science
Institute for granting Discretionary Director's Time, and
Dr. Mark Hindmarsh for his fruitful comments.

M.V. Sazhin acknowledges the VSTceN-INAF for hospitality and
financial support, and the financial support of RFFI grant
04-02-17288. O.S. Khovanskaya acknowledges the INFN-Napoli and the
Department of Physical Sciences at the University Federico II for
financial support, as well as the financial support of the grants:
of President of RF "YS-1418.2005.2" and INTAS Ref. Nr.
05-109-4793. This research was funded by the Italian Ministry
MIUR through a 2004 PRIN grant (2004020323\_006) and by Regione Campania through a L41 grant.
N.A. Grogin acknowledges financial support from HST Grant GO-10715-A.

\clearpage

\newpage

\newpage

\label{lastpage}


\begin{thebibliography}{99}


\bibitem[\protect\citeauthoryear{Agol et al.}{2006}]{ago06}
Agol, E., Hogan, C.~J., \& Plotkin, R.~M.\ 2006, Phys.Rev. D, 73, 087302 


\bibitem[\protect\citeauthoryear{Alcal{\'a} et al.}{2004}]{alc04}
Alcal\'a J.M., et al., 2004, A\&A, 428, 339

\bibitem[\protect\citeauthoryear{Allen \& Shellard}{1990}]{all}
Allen B. and Shellard E.P.S., 1990, Phys.Rev.Lett., 64, 119.

\bibitem[\protect\citeauthoryear{Bennett \& Bouchet}{1990}]{Bennet90} 
Bennett, D.~P., \& Bouchet, F.~R.\ 1990, Phys.Rev. D, 41, 2408 

\bibitem[\protect\citeauthoryear{Bernardeau \& Uzan}{2000}]{string3}
Bernardeau F., Uzan J.-P., 2001, Phys.Rev D, 63, 023004,
023005

\bibitem[\protect\citeauthoryear{Bevis et al.}{2004}]{bev04}
Bevis, N., Hindmarsh, M., \& Kunz, M.\ 2004, Phys.Rev. D, 70, 043508 

\bibitem[\protect\citeauthoryear{Bevis et al.}{2006}]{bev06}
Bevis N., Hindmarsh M., Kunz M., \& Urrestilla J., 2006, astro-ph/0605018.

\bibitem[\protect\citeauthoryear{Capaccioli et al.}{2002}]{cap1}
Capaccioli M., Alcal\'a J.M., Radovich M., Silvotti R., Ortiz
P.F., et al., 2002, Proc. SPIE 4836,  pp.

\bibitem[\protect\citeauthoryear{Carretero et al.}{2006}]{carret}Carretero et al. 2006, astro-ph/0608012

\bibitem[\protect\citeauthoryear{Casertano et al.}{2000}]{casertano00} Casertano, S., et
al., 2000, AJ, 120, 2747

\bibitem[\protect\citeauthoryear{Chiba \& Yoshi}{1999}]{chi99}
Chiba M., Yoshi Yu., 1999, Ap.J., 510, 42

\bibitem[\protect\citeauthoryear{Copeland et al.}{2004}]{cop04}
Copeland E.J., Myers R.C., Polchinski J., 2004, JHEP, 06:013.


\bibitem[\protect\citeauthoryear{Davis \& Kibble}{2005}]{dav05}
Davis A.-C.,  Kibble T.W.B., 2005, hep-th/0505050

\bibitem[\protect\citeauthoryear{Damour and Vilenkin}{2004}]{dam04}
Damour T. and Vilenkin A., hep-th/0410222.

\bibitem[\protect\citeauthoryear{Fruchter \& Hook}{2002}]{fruch02} Fruchter, A.~S., \&
Hook, R.~N., 2002, PASP, 114, 144

\bibitem[\protect\citeauthoryear{Fukugita et al.}{1992}]{fuk92}
Fukugita, M., Futamase, T., Kasai., Turner E.L., 1992, Ap.J., 393,
3

\bibitem[\protect\citeauthoryear{Gardner et al.}{1996}]{cou}
Gardner J.P., Sharples R.M., Carraso B.E., Frenk C.S., 1996, MNRAS,
282, L1

\bibitem[\protect\citeauthoryear{Gardner}{1998}]{gar98}
Gardner J., 1998, PASP, 110, 291

\bibitem[\protect\citeauthoryear{de Vaucouleurs}{1953}]{GV_53}
de Vaucouleurs, G. 1953, MNRAS, 113, 134

\bibitem[\protect\citeauthoryear{Shanks et al.,}{1998}]{sha98}
Shanks T., Metcalfe N., Fong R., et al., 1998, in: The Young
Universe, ASP Conference Ser., Eds. S.D'Odorico, A.Fontana,
E.Gialongo, v.146, p.102.

\bibitem[\protect\citeauthoryear{Gardner and Satyapal}{2000}]{gar00}
Gardner J., Satyapal S., Astron.J., 2000, 119, 2589

\bibitem[\protect\citeauthoryear{Hindmarsh}{1990}]{string2}
Hindmarsh. A.,in "The Formation and Evolution of Cosmic Strings", ed.
 G.Gibbons, S.W.Hawking \& T.Vachaspathi.  Cambridge
Univ.Press., Cambridge, 1990.


\bibitem[\protect\citeauthoryear{Huterer \& Vachaspati}{2003}]{string5}
Huterer, D., Vachaspati, T., 2003, preprint. astro-ph/0305006.

\bibitem[\protect\citeauthoryear{Kibble}{1976}]{kib76}
Kibble T.W.B., 1976, J.Phys.A:Math \& Gen. v.9, 1387

\bibitem[\protect\citeauthoryear{Kochanek}{2002}]{kee}
Kochanek et al., 2002, http://cfa-www.harvard.edu/castles/

\bibitem[\protect\citeauthoryear{Kochanek}{1993}]{koc93}
Kochanek C.S., 1993, MNRAS, 261, 453

\bibitem[\protect\citeauthoryear{Koekemoer et al.}{2002}]{koe02}
Koekemoer, A.M., Fruchter, A.S., Hook, R., Hack, W. 2002 HST
Calibration Workshop, 337.

\bibitem[\protect\citeauthoryear{Kummel \& Wagner}{2001}]{couR}
Kummel M.W., Wagner S.J., 2001, astro-ph/0102036.

\bibitem[\protect\citeauthoryear{Laix \& Vachaspati}{1996}]{string4}
de Laix, A.A., Vachaspati, T., 1996, Phys.Rev. D 54, 4780, 1996.

\bibitem[\protect\citeauthoryear{Majumdar}{2005}]{maj05}
Majumdar M., hep-th/0512065.

\bibitem[\protect\citeauthoryear{Ofek et al.}{2003}]{ofe03}
Ofek E.O., Rix H-W., Maoz D., 2003, MNRAS, 343, 639

\bibitem[\protect\citeauthoryear{Polchinski and Rocha}{2006}]{J.Polchinski}
Polchinski, J. \& Rocha, J.V., hep-ph/0606205

\bibitem[\protect\citeauthoryear{Ringeval et al.}{2005}]{Rigv05}
Ringeval C., Sakellariadou M. \& Bouchet F., 2005, astro-ph/0511646.

\bibitem[\protect\citeauthoryear{Sazhin et al.}{2003}]{csl1}
Sazhin, M., et al., 2003, MNRAS, 343, 353

\bibitem[\protect\citeauthoryear{Sazhin et al.}{2005}]{saz05}
Sazhin, M., Capaccioli, M., Longo, G., Paolillo, M., \& Khovanskaya, O.\ 2006, ApJ, 636, L5

\bibitem[\protect\citeauthoryear{Sazhin et al.}{2006}]{saz06}
Sazhin, M., et al., 2006, astro-ph/0601494

\bibitem[\protect\citeauthoryear{Schneider, Ehlers, Falco}{1992}]{sch92}
Schneider P., Ehlers J., Falco E.E., 1992, Gravitational Lenses,
Springer, Heidelberg.

\bibitem[\protect\citeauthoryear{Shirasaki, Mizumoto, Ohishi}{2004}]{shi04}
Shirasaki, Y., Mizumoto, Y., Ohishi, M. et al. ASP Conference
Series, vol. 314, 2004.

\bibitem[\protect\citeauthoryear{Shlaer and Tye}{2005}]{shl05}
Shlaer B., Tye S.-H. Henry, hep-th/0502242.


\bibitem[\protect\citeauthoryear{Thomson et al.}{1999}]{hubbledf1}
Thomson, R.I., Storrie -Lombardi, L.J., Weymann, R., et al., 1999,
Astron.J., 117, 17.

\bibitem[\protect\citeauthoryear{Tye et al.}{2005}]{tye05}
Tye S.-H.Henry, Wasserman Ira and Mark Wyman, astr-ph/0503506.

\bibitem[\protect\citeauthoryear{Vilenkin}{1981}]{vil}
Vilenkin A., 1981, Phys.Rev. D, 23, 852

\bibitem[\protect\citeauthoryear{Vilenkin}{1984}]{vil84}
Vilenkin A., 1984, ApJ, 289, L51.

\bibitem[\protect\citeauthoryear{Vilenkin}{1986}]{vil86}
Vilenkin A., 1986, Nature, 322, 613.


\bibitem[\protect\citeauthoryear{Vilenkin, Shellard}{1994}]{string1}
Vilenkin A., Shellard E.P.S, 1994, Cosmic strings and other
topological defects. Cambridge Univ.Press., Cambridge.

\bibitem[\protect\citeauthoryear{Vincent et al.}{1998}]{vin98} 
Vincent, G., Antunes, N.~D., \& Hindmarsh, M.\ 1998, Phys. Rev. Letters, 80, 2277 

\bibitem[\protect\citeauthoryear{Williams et al.}{1996}]{Williams96}
Williams, R. E.; Blacker, B.; Dickinson, M. Astron.J.,
1996, v.112, p.1335.

\bibitem[\protect\citeauthoryear{Zakharov and Sazhin}{1998}]{saz98}
Zakharov A.F., Sazhin, M.V., 1998, PHYS-USP, 41(10), 945.


\bibitem[\protect\citeauthoryear{Zeldovich}{1980}]{zel}
Zeldovich, Ya.B., 1980, MNRAS, 192, 663

\end{thebibliography}
\end{document}